\documentclass[final]{aipproc}
\usepackage{graphics}
\layoutstyle{6x9}
\begin{document}
\title{Field-doping of C$_{60}$ crystals: A view from theory}
\author{Erik Koch}{address={Max-Planck-Institut f\"ur Festk\"orperforschung,
                   Heisenbergstra\ss e 1, D-70569 Stuttgart}}
\author{Olle Gunnarsson}{address={Max-Planck-Institut f\"ur
        Festk\"orperforschung, Heisenbergstra\ss e 1, D-70569 Stuttgart}}
\author{Samuel Wehrli}{address={Theoretische Physik, ETH-H\"onggerberg,
                       CH-8093 Z\"urich}}
\author{Manfred Sigrist}{address={Theoretische Physik, ETH-H\"onggerberg,
                         CH-8093 Z\"urich}}

\begin{abstract}
The proposal of using the field-effect for doping organic crystals has raised
enormous interest. To assess the feasibility of such an approach, we 
investigate the effect of a strong electric field on the electronic structure
of C$_{60}$ crystals. Calculating the polarization of the molecules and the
splittings of the molecular levels as a function of the external field, we
determine up to what field-strengths the electronic structure of C$_{60}$
stays essentially unchanged, so that one can speak of field-effect doping, in
the sense of putting charge carriers into otherwise unchanged states.
Beyond these field strengths, the electronic structure changes so much, that 
on can no longer speak of a doped system.
In addition, we address the question of a metal-insulator transition at 
integer dopings and briefly review proposed mechanisms for explaining an
increase of the superconducting transition temperature in field-doped
C$_{60}$ that is intercalated with haloform molecules.
\end{abstract}
\maketitle

\noindent
The doped fullerenes are materials with very interesting properties.
Alkali doped C$_{60}$ with three alkalis per molecule has, e.g., turned out 
to be metallic, though close to a Mott transition, and superconducting.
A problem is, however, that different doping levels can only be realized
by preparing separate crystals. Moreover, because of the strong 
electronegativity of C$_{60}$, no hole-doping has been achieved.
In this context the proposal of using a field-effect transistor for doping
pristine C$_{60}$ crystals has raised much interest, in particular since 
such an approach should allow us to continuously change the doping by simply 
changing the voltage applied to the gate electrode. Sadly the reports of such 
field-doping and of spectacular values for the superconducting transition 
temperatures in such devices \cite{elecdoped,holedoped,latticeexp} have been 
withdrawn \cite{retract} after an investigation showed that the publications 
were based on fraudulent data \cite{fraud}.
Nevertheless, it is an open question whether field-doping of C$_{60}$
crystals could be achieved in principle. 
In the following we address several aspects of this question.
%This is the question we address in the following. 

\section{Effect of a strong electric field}
Reaching substantial charging (of the order of $n$ electrons per C$_{60}$
molecule) in a field-effect device requires enormous electric fields. 
As the induced charge is basically restricted to one monolayer of C$_{60}$ 
\cite{wehrli}, a rough estimate can be obtained from simple electrostatics: 
For neutrality the charge on the gate must equal that on the monolayer,
hence the field originating from the gate electrode is given by
$E_{\mathrm{gate}}=2\pi\,n/A_{\mathrm{mol}}$, where $A_{\mathrm{mol}}$ is
the area per C$_{60}$ molecule in the monolayer and $n$ is the number of
induced electron charges per molecule. Thus the external field is about 
1~V/\AA\ per induced charge, corresponding to a voltage drop of about 10~V 
across the molecule. In such a strong external field the C$_{60}$ molecules are 
strongly polarized. Nevertheless, we find that their response is still in the 
linear regime. 
Furthermore, in the charged monolayer, the field experienced by a molecule 
is screened by the polarization of the neighboring molecules. Taking this 
into account, we find that the field is reduced by about a factor of two. 
Calculating the splitting of the molecular levels in this screened homogeneous 
field, we find a quadratic Stark effect, with the splitting of both, the 
$t_{1u}$ and the $h_u$ level, becoming of the order of the band-width for 
a field corresponding to an induced charge of three to four charges per 
molecule. This seems to be consistent with the typical doping levels that
had been reported.

For a more realistic description of the electrostatics in the field-effect
device, we have, however, to go beyond considering only a homogeneous field.
Given the spherical shape of C$_{60}$, the natural approach is via a multipole
expansion \cite{multipole}: We choose an external field and the corresponding
induced charge per molecule. We determine the multipole expansion of the field 
generated by all other molecules about the molecule centered at the origin.
Using the linear response of a C$_{60}$ molecule to multipole fields 
(calculated {\em ab initio}), we determine the new charge distribution on 
the molecules and repeat the procedure until self-consistency is reached.

\begin{figure}
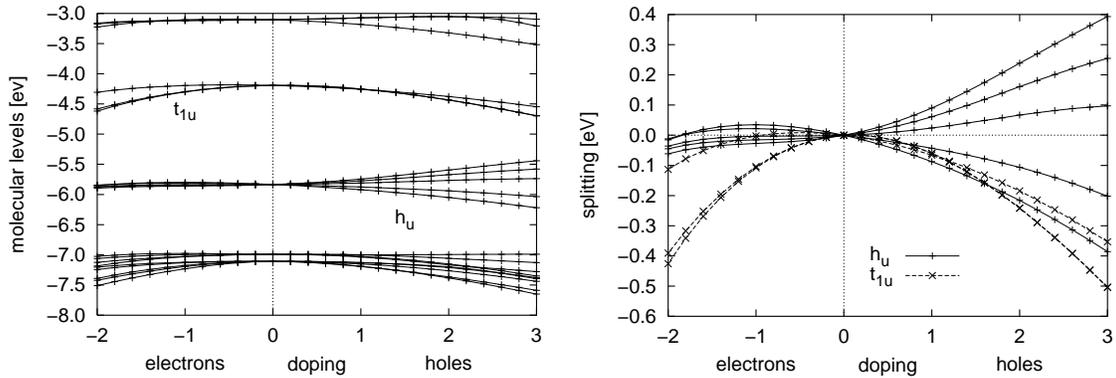

 \centerline{
 \resizebox{0.48\textwidth}{!}{\rotatebox{270}{\includegraphics{C2vveffsqul.epsi}}}
 \hspace{0.02\textwidth}
 \resizebox{0.48\textwidth}{!}{\rotatebox{270}{\includegraphics{C2vveffsqus.epsi}}}}

 \vspace{1ex}
 \label{split}
 \caption{Splitting of the molecular levels in the self-consistent multipole
          field ($l\le2$) for a (001) monolayer (square lattice) of C$_{60}$
          molecules oriented such that one of their two-fold axes points in
          the direction of the external electric field (perpendicular to the
          layer).}
\end{figure}

Figure \ref{split} shows the splitting of the molecular levels in the
self-consistent multipole field for different doping levels. While for an
external homogeneous field ($l=1$ multipole) the splitting is independent 
of the direction of the field, including the effect of the induced charge on 
the neighboring molecules breaks this symmetry. Surprisingly, the 
asymmetry in the splitting is quite strong, even though the fields that
break the symmetry are fairly weak compared to the homogeneous field.
This is because the multipole potentials with even $l$ give
rise to a {\em linear} Stark effect, which changes sign with the external
potential and which gives rise to a strong splitting even for weak fields.
In addition it turns out \cite{multipole} that the splitting due to the
$l=1$ and $l=2$ potentials add or subtract, depending on the sign of the
external field: When inducing electrons they add for the $t_{1u}$ level
and almost cancel for the $h_u$, while when inducing holes the situation
is reversed. I.e., when a molecular level is filled, the splitting is
substantially enhanced. It reaches the order of the band width when inducing 
about two electrons, or somewhat more than two holes per molecule. Beyond these
fillings the effect of the Stark splitting on the electronic structure
of the C$_{60}$ monolayer that carries the induced charge will clearly
be very large, and one can definitely no longer speak of doping.

\section{HOMO-antiscreening}
\begin{figure}
 \begin{minipage}{\textwidth}
  \centering
  \resizebox{0.35\textwidth}{!}{\includegraphics{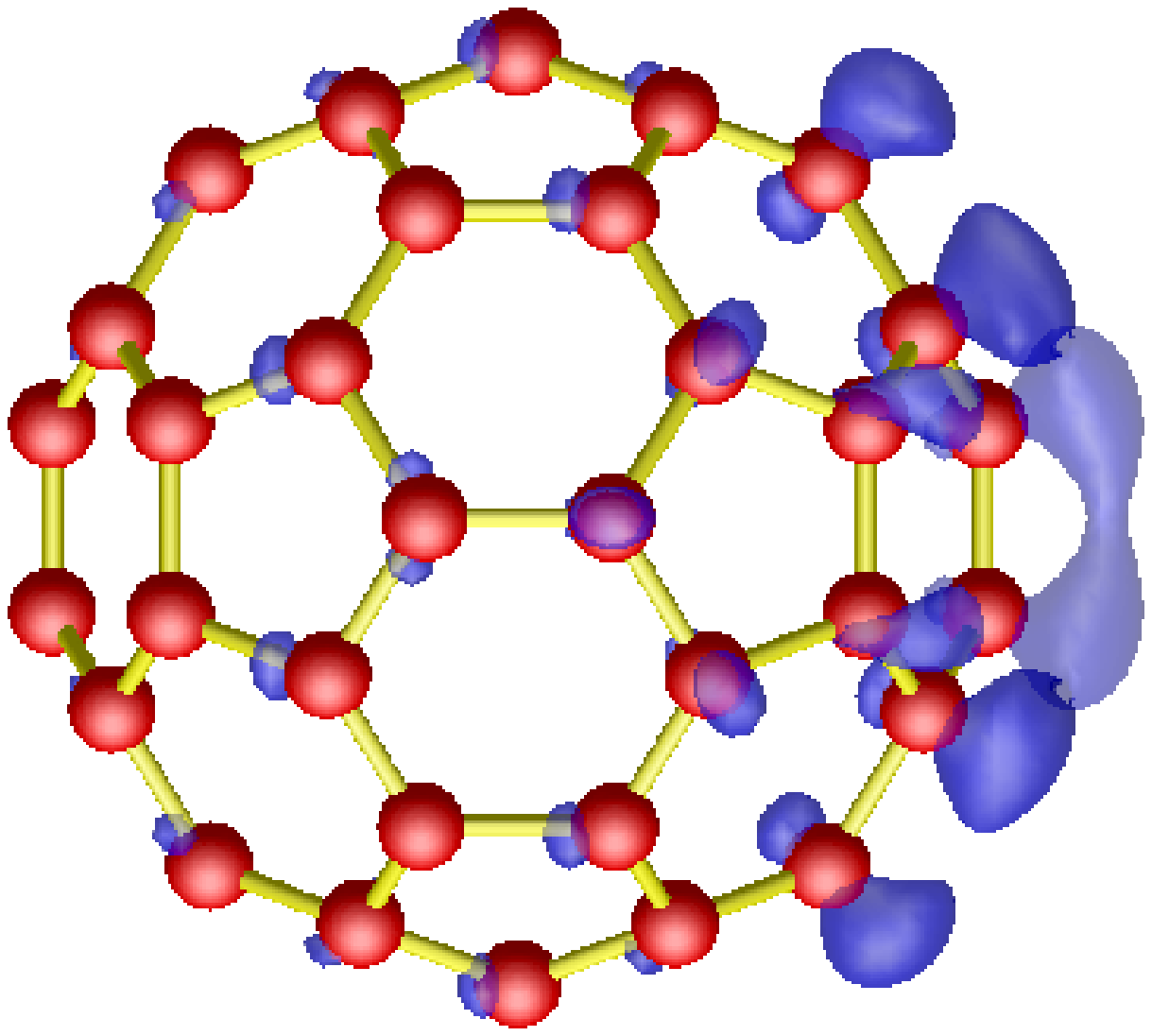}}
  \hspace{8ex}
  \resizebox{0.35\textwidth}{!}{\includegraphics{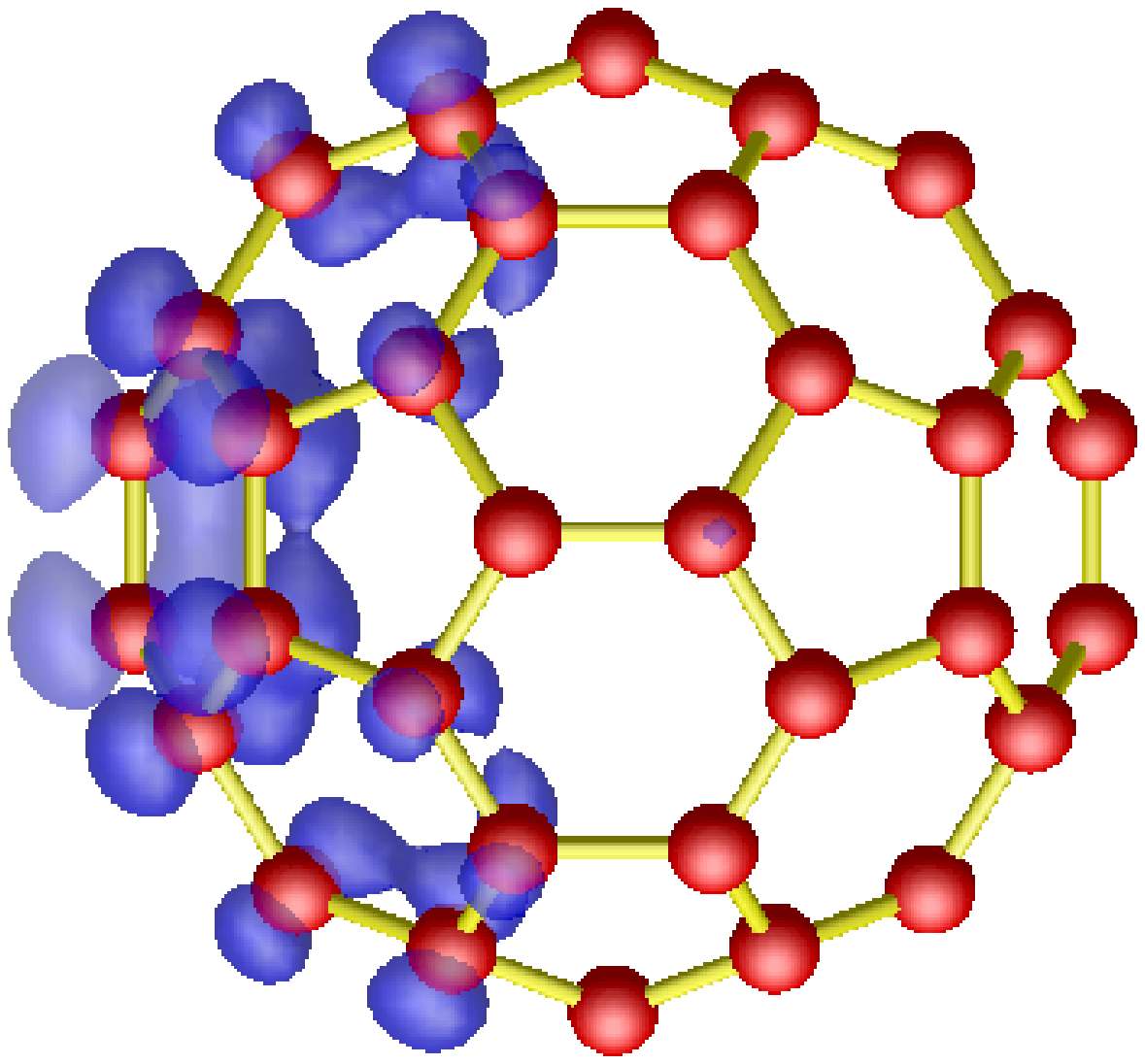}}\\[1ex]
  {\bf total charge\hspace{26ex}$h_u$ orbital}\\[1ex]
  11.2 $e\,a_0$\hspace{4ex}dipole moment\hspace{12ex}--8.7 $e\,a_0$
 \end{minipage}

 \vspace{1ex}
 \label{antis}
 \caption{Polarization of a C$_{60}$ molecule in a homogeneous external
          electric field of 0.02~a.u. ($\approx$~1~V/\AA). The change in
          charge density compared to the field-free case is indicated by
          the $\Delta\rho$-isosurface at 0.0020/$a_0^3$. It turns out
          that the dipole moment for the $h_u$ (HOMO) charge density
          is of the same order of magnitude as that of the total charge ---
          {\it but of the opposite sign.}} 
\end{figure}

As we have seen, the charge density of the C$_{60}$ molecule is strongly
polarized in an electric field, and one might expect that the main contribution
comes from the polarization of the highest molecular orbital (HOMO).
Calculating the change in the HOMO charge density, 
% $\Delta\rho_{HOMO}=\sum_{n\in HOMO}\left(\langle\Psi_n|\Psi_n\rangle
%                                    -\langle\Psi_n^0|\Psi_n^0\rangle\right)$,
we find, however, that the dipole moment of the HOMO charge density is of
the same order of magnitude as that of the total charge -- {\it but of the 
opposite sign} (see Fig.~\ref{antis}). This surprising result can be 
understood, e.g., in terms of perturbation theory: Expanding the wave function 
to first order in the external field $V=E_z\,z$ and calculating the dipole 
moment $p=e\,z$, we find that the leading term is given by
a sum over the matrix element, squared, with all unperturbed molecular orbitals
of different parity divided by the characteristic energy denominator.
Hence the main contribution comes from energetically close-by levels, and
the sign of their contribution is determined by whether they are energetically
above or below the level under consideration. For the HOMO in a molecule with
a large HOMO-LUMO gap this means that the contributions mainly come from 
the molecular levels below -- implying antiscreening. We thus see that
HOMO-antiscreening should be quite general for molecules with large HOMO-LUMO
gap, and, in fact, it can also be found, e.g., in the series of polyacenes:
benzene, naphthalene, anthracene, tetracene, and pentacene. 

\section{Mott transition}
\begin{figure}
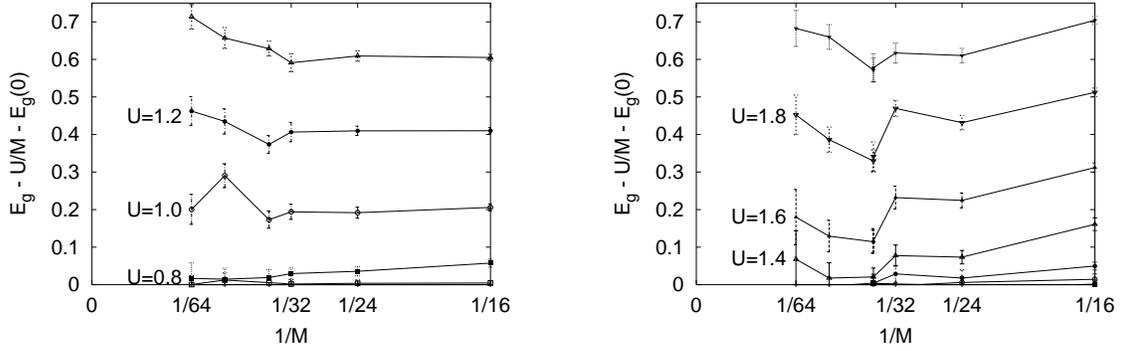

 \centerline{
 \resizebox{0.45\textwidth}{!}{\rotatebox{270}{\includegraphics{Egap.t1u.epsi}}}
 \hspace{0.08\textwidth}
 \resizebox{0.45\textwidth}{!}{\rotatebox{270}{\includegraphics{Egap.hu.epsi}}}}
 \label{Egap}
 \caption{Gap $E_g=E(N\!-\!1)-2E(N)+E(N\!+\!1)$ with finite-size
          correction $E_g-E_g(U\!=\!0)-U/M$ as calculated by quantum
          Monte Carlo for the $t_{1u}$ and the $h_u$ band in the
          (111)-plane of the Pa$\bar{3}$ structure. For the half-filled
          $t_{1u}$-band the gap opens between $U_c=0.8\ldots1.2$~eV, for
          the half-filled $h_u$-band between $U_c=1.2\ldots1.6$~eV.}
\end{figure}

Since the bands in the fullerenes are narrow, while the Coulomb repulsion
between two electrons on the same molecule is sizable, the doped fullerenes
show effects of strong correlation. It is, e.g., only due to orbital degeneracy 
that A$_3$C$_{60}$ is metallic and not a Mott insulator \cite{c60mott}.
In field-doped fullerenes the electrons are restricted to a monolayer
\cite{wehrli}. Hence the number of nearest neighbors to which an electron
can hop is reduced and the bands are even more narrow.
It is therefore expected that the Mott transition occurs at critical values
of the Coulomb interaction $U$ below those found in the bulk. To determine
the transition point, we have performed quantum Monte Carlo calculations
\cite{filling} for a doped (111)-layer (without Stark splitting) and find that 
the Mott transition occurs between $U_c=0.8-1.2$~eV for doping with three 
electrons, and $U_c=1.2-1.6$~eV for the half-filled $h_u$ band 
(see Fig.~\ref{Egap}). 
For integer dopings other than half-filling the transition is expected to 
occur for even smaller values of $U$ \cite{filling}. Furthermore, the splitting
of the molecular levels in the electric field should weaken the effect of the 
degeneracy on the Mott transition and lead to still smaller values of $U_c$ 
\cite{manini}. In particular for the $h_u$ orbital, the strong electron-phonon
coupling might lead to a further reduction of $U_c$ \cite{JThund}.
One has, however, to keep in mind that the Coulomb interaction 
$U$ depends on the environment of the molecule. Screening due to the 
polarization of the neighboring molecules is, e.g., responsible for a large 
reduction of $U$ in the crystal as compared to the value for an isolated 
molecule \cite{c60U}. Likewise, it is to be expected that for a molecule in the
monolayer next to the gate dielectric, $U$ might be substantially changed from 
the bulk value. This effect is, however, hard to quantify without knowing the 
microscopic structure of the oxide-C$_{60}$ interface. 
Nevertheless it seems likely that field-doped C$_{60}$ should be insulating
at integer fillings.

\section{Enhancement of transition temperature}
In A$_3$C$_{60}$ the superconducting transition temperature $T_c$ increases
with increasing lattice constant, i.e., with increasing density of states
at the Fermi level \cite{rmp}. It is therefore natural to try the
same for field-doped C$_{60}$. The simplest way to increase the 
distance of the molecules in the conducting monolayer is to apply uniaxial
stress \cite{uniax}. An alternative approach is the intercalation of the
crystal with inert molecules. In fact, for field-doped C$_{60}$ 
intercalated with haloform molecules spectacularly increased transition
temperatures have been reported \cite{latticeexp,retract}. A subsequent
analysis of the lattice structure of these crystals revealed, however,
that the lattice is mainly expanded perpendicular to the conducting layer,
and that the density of states in the doped layer shows no correlation with
the reported $T_c$ \cite{jansen}. Therefore, the additional coupling to the
vibrations of the haloform molecules has been proposed as an alternative 
explanation of the enhancement of the transition temperature \cite{bill}.
It has, however, turned out that such a coupling is very small and, for
the two-fold degenerate modes, is even excluded by symmetry \cite{CHX3}.
The reported enhancement of $T_c$ in haloform intercalated C$_{60}$ is
therefore not understood.

\bibliographystyle{aipproc}

\end{document}